\documentclass[a4paper,article,twocolumn,showkeys,aps,prl,twoside,amsmath,amssymb]{revtex4-2}
\usepackage{natbib}
\usepackage[pdftex]{hyperref}
\usepackage{graphicx}
\usepackage{lineno}
\usepackage{xcolor}
\usepackage{float}
\usepackage{siunitx}
\usepackage{booktabs}
\usepackage{fancyvrb}
\usepackage{lettrine}

\hypersetup{
	colorlinks,
	linkcolor={blue},
	citecolor={red!50!black},
	urlcolor={blue!80!black},
	frenchlinks=true,
	pdfauthor={Dr. Jakob Wierzbowski},
	pdftitle={On Prism Cross-Dispersers - Modeling Échelle Spectrograms},
	pdfsubject={Modelling prisms the right way - with QtYETI.},
	pdfkeywords={Prism; Echelle; Cross-Disperser; Snell's law; Sellmeier; QtYETI},
	pdfproducer={LateΧ},
	pdfcreator={pdflatex}
}

\newcommand{\figref}[2]{Figure~\textbf{\ref{#1}#2}}

\begin{document}

\title{On Prism Cross-Dispersers - Modelling \'{E}chelle Spec\-tro\-grams}
\author{Jakob Wierzbowski}
\email{jakob.wierzbowski@wsi.tum.de}
\affiliation{Walter Schottky Institut der TU M\"{u}nchen, Am Coulombwall 4, 85748 Garching bei M\"{u}nchen}
\author{Bernd Bitnar}
\altaffiliation{Contributed equally to this work}
\noaffiliation
\author{Siegfried Hold}
\altaffiliation{Contributed equally to this work}
\noaffiliation

\begin{abstract}
\noindent In this paper, we elaborate on correctly predicting \'{E}chelle spec\-tro\-grams by employing the fully three-dimensional representation of Snell’s law to model the effects of prisms as cross-dispersers in \'{E}chelle spectrographs.
We find that it is not sufficient to simply apply the frequently used tri\-go\-no\-met\-ric prism dispersion equation to describe recorded spectra.
This vector equation approach is not limited to a single dispersive element when modelling multi-prism cross-disperser configurations.
Our results help to understand the main levers in an \'{E}chelle spectrograph as well as contribute to auto-calibration algorithms for minimizing calibration efforts in daily operation.
\end{abstract}

\maketitle

\section{Introduction}
\lettrine[lines=2, findent=2pt, nindent=0pt]{P}{risms} are key parts in compact \'{E}chelle spectrographs which, in turn, are widely used in astrospectroscopy \cite{bib:Schroeder1971,bib:Hamilton1987,bib:Orfeus1988,bib:Pepsi2018}.
The advantage of prisms as dispersive elements is their high optical efficiency and transmittance in contrast to optical gratings.
Unlike gratings, prisms do not diffract incident light into multiple orders but rather refract it into specific directions.
To harvest the largest amount of light while covering a large spectral band, a common approach is to first disperse collimated light on a blazed, reflective optical grating with a low groove density on the order of d = 30 to 80 $\SI{}{\milli\meter}^{-1}$ and subsequently cross-disperse the diffracted light via a prism before focusing it down onto a detector using a lens.
The low groove density helps to generate multiple overlapping orders ($m = 20\ \text{to}\ 60$), e.g. in the visible range.
%
%
At the same time, \'{E}chelle gratings are specifically designed with comparably high blaze angles $(\theta_B > \SI{60}{\degree})$.
Using them close to the Littrow condition $\alpha \approx \beta$ for highest optical efficiency, \'{E}chelle grating typically exhibit a large geometrical width.
This increases the number of illuminated grooves.
Both design parameters in turn help to achieve a high resolution $R = \lambda/\Delta\lambda$.
An elaborate discussion on grating physics and blazed gratings is beyond the scope of this work.
Comprehensive and beginner-friendly books on grating physics and astrospectroscopy with \'{E}chelles can be found in \cite{bib:Palmer2014} and \cite{bib:Schanne2018}, respectively.\newline
Nevertheless, the dispersion relation of a diffraction grating is accurately described by the heuristic equation:
\begin{equation}
	m\cdot\lambda=d(\sin{\alpha} + \sin{\beta})\cos{\gamma}\label{eq:grating},
\end{equation}
where m is the order number, $\lambda$ the wavelength, d the grating constant, $\alpha$ and $\beta$ the in- and outgoing angles, respectively.
$\gamma$ is the relative angle with respect to the plane of incidence, thus, perpendicular to $\alpha$ and $\beta$.

\begin{figure*}[t!]
	\includegraphics[width=1\textwidth,keepaspectratio]{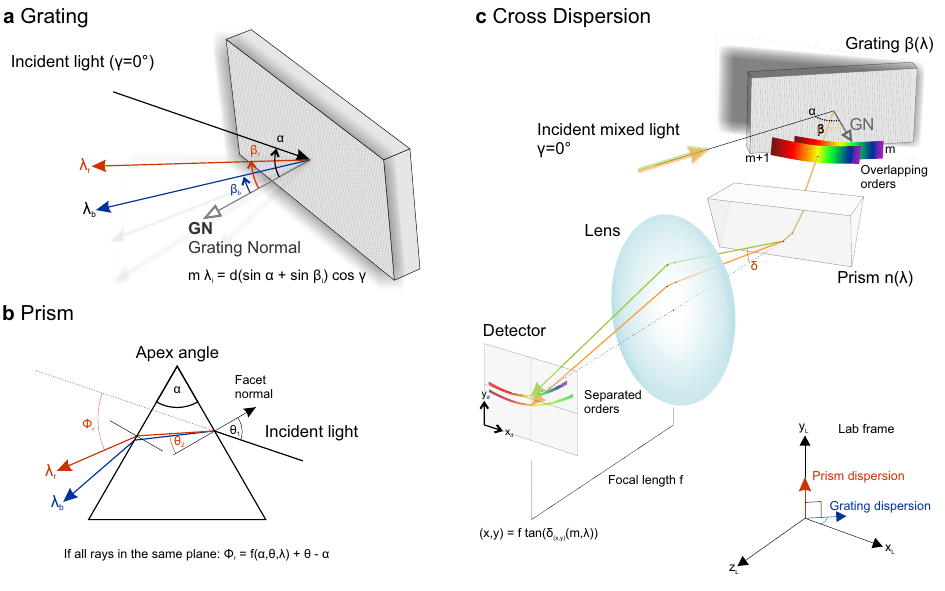}
	\caption{Principles of dispersion. \textbf{a}. Diffraction Grating \textbf{b}. Dispersion prism. \textbf{c}. Cross-Disperser principle. Light incident on a grating at angles $\left(\alpha,\gamma\right)$ is diffracted into the direction $\beta$ according to the grating equation. The resulting and overlapping orders $m_i$ travel through a prism whose dispersion direction is perpendicular relative to the grating (red and blue arrows). This principle separates overlapping spectral orders into unique positions on the detector.}
	\label{fig:figure1}
\end{figure*}

Employing equation (\ref{eq:grating}), we use the standard convention of measuring angles \textit{from} the grating normal $\mathbf{GN}$ \textit{towards} the in- and outgoing beam, respectively \cite{bib:Palmer2014}.
Angles that are measured in clockwise/counter-clockwise direction are negative/positive.
\figref{fig:figure1}{a} summarizes this convention. Here, the incident angle $\alpha$ and the outgoing angles $\beta_b$ and $\beta_r$ are negative since they are measured in clockwise direction \textit{from} the grating normal $\mathbf{GN}$.
In stark contrast to gratings, which exploit interference at periodic structures to diffract light, prisms on the other hand disperse light by refraction driven by a wavelength dependent refractive index within media \cite{bib:griffiths2013introduction} such as glasses.

\section{Modelling \'{E}chelles}
\subsection{Geometric prism equation}
\noindent \figref{fig:figure1}{b} shows how incident light entering a prism at an angle $\theta_1$ is refracted according to Snell’s law $n_1\cdot \sin{\theta_1}=n_2 \cdot \sin{\theta_2}$, with $n_1$ and $n_2$ being the refractive indices of e.g. air and the prism medium (e.g. glass, $n > 1$), respectively.
Both angles, $\theta_1$ and $\theta_2$ are measured relative to the surface normal describing the optical interface.
The same principle applies to light exiting the prism.
Using Snell’s law, trigonometric relations as well as setting $n_1=1$, we can approximate the total refraction angle $\Phi\left(\lambda\right)$ to be:
\begin{align}
	\Phi\left(\lambda\right) &= \arcsin\left[\ n_{\lambda}\cdot\sin \alpha-\arcsin(\sin{\theta/n_{\lambda}})\ \right]
	\label{eq:prismgeo}\\
		\nonumber
		&+ \theta - \alpha,
\end{align}
\noindent where $\alpha$ is the apex angle of the prism, $\theta = \theta_1$ is the angle between the prisms surface normal and the incoming beam as well as $n_{\lambda} = n_2$ being the wavelength dependent refractive index of the prism.
For glass prisms, the refractive index is best described by the Sellmeier equation~\cite{bib:Sellmeier1871} (see supplementary material).

Please note, equation (\ref{eq:prismgeo}) is a strong simplification, which will turn out to deliver incorrect results when trying to predict \'{E}chelle spectra.
It only applies when incoming and outgoing light lies in the same plane as the facet's surface normal vector.

\subsection{Cross-dispersion with prisms}
\noindent \'{E}chelle spectrographs often use diffraction gratings with a low number of grooves per mm as main dispersing element in combination with, either, a second grating oriented perpendicularly to the first one, or a \emph{prism}.
Due to the low groove density, the diffraction orders tend to strongly overlap for $|m| > 20$ in the visible range \cite{bib:Palmer2014,bib:Schanne2018}.
Therefore, disentangling the light for different orders becomes a necessity.
\figref{fig:figure1}{c} shows a typical \'{E}chelle cross-disperser configuration, where the second dispersion element is a prism.
A collimated beam of light that is parallel to the laboratory xz-plane ($\gamma = \SI{0}{\degree}$) hits the surface of a reflective diffraction grating at a fixed angle $\alpha$.
The grating disperses the light in the xz-plane into multiple and most importantly overlapping orders $m_i$.
Here, $\alpha$ and the outgoing angle $\beta$ are negative with respect to the grating normal $\mathbf{GN}$.
Taking a step back to the grating equation (\ref{eq:grating}), it is worth noting that the left side of the equation is changing, both, with $\lambda$ and $m$.
For a fixed observation angle $\beta = const.$, the right-hand side of the equation becomes a set of constants.

Suppose $\beta = \beta_o = \beta_g = const.$, it is then possible to find two wavelengths $\lambda_o$ and $\lambda_g$, such that:\newline
\begin{equation}
	m_o\lambda_o=m_g\lambda_g,\quad m_{o,g}\in \mathbb{Z}\setminus\left\{0\right\}\label{eq:gratingsamedirection}
\end{equation}
\noindent Thus, equation (\ref{eq:gratingsamedirection}) dictates that two wavelengths from different orders satisfying this relation will be diffracted into the same direction.
The two arrows (orange and green) in \figref{fig:figure1}{c} represent two beams of light with wavelengths $\lambda_o$ and $\lambda_g$, respectively.
Both beams are diffracted into the same direction depicted by the angle $\beta$ whilst belonging to neighbouring diffraction orders $m_o$ and $m_g = m_o + 1$.
Upon exiting the prism at the second facet, both light rays will leave the prism under different angles, as previously explained in \figref{fig:figure1}{b}.
Subsequently, an objective lens focuses these light beams onto a detector.
The position $y_d$ on the detector is determined by the angle $\delta$ of the green/orange light beam relative to the optical axis of the lens and its focal length f, thus, $y_d=f\cdot\tan{\delta}$.
Since $\delta_o \neq \delta_g$, both light rays will be focused onto different points $(x_d,y_{d,o})$ and $(x_d,y_{d,g})$ on the detector.
Applying this principle to all available light beams within the field of view of the prism and lens, we are now able to separate all physical orders.
\begin{figure*}[t!]
	\centering
	\includegraphics[width=1\textwidth,keepaspectratio]{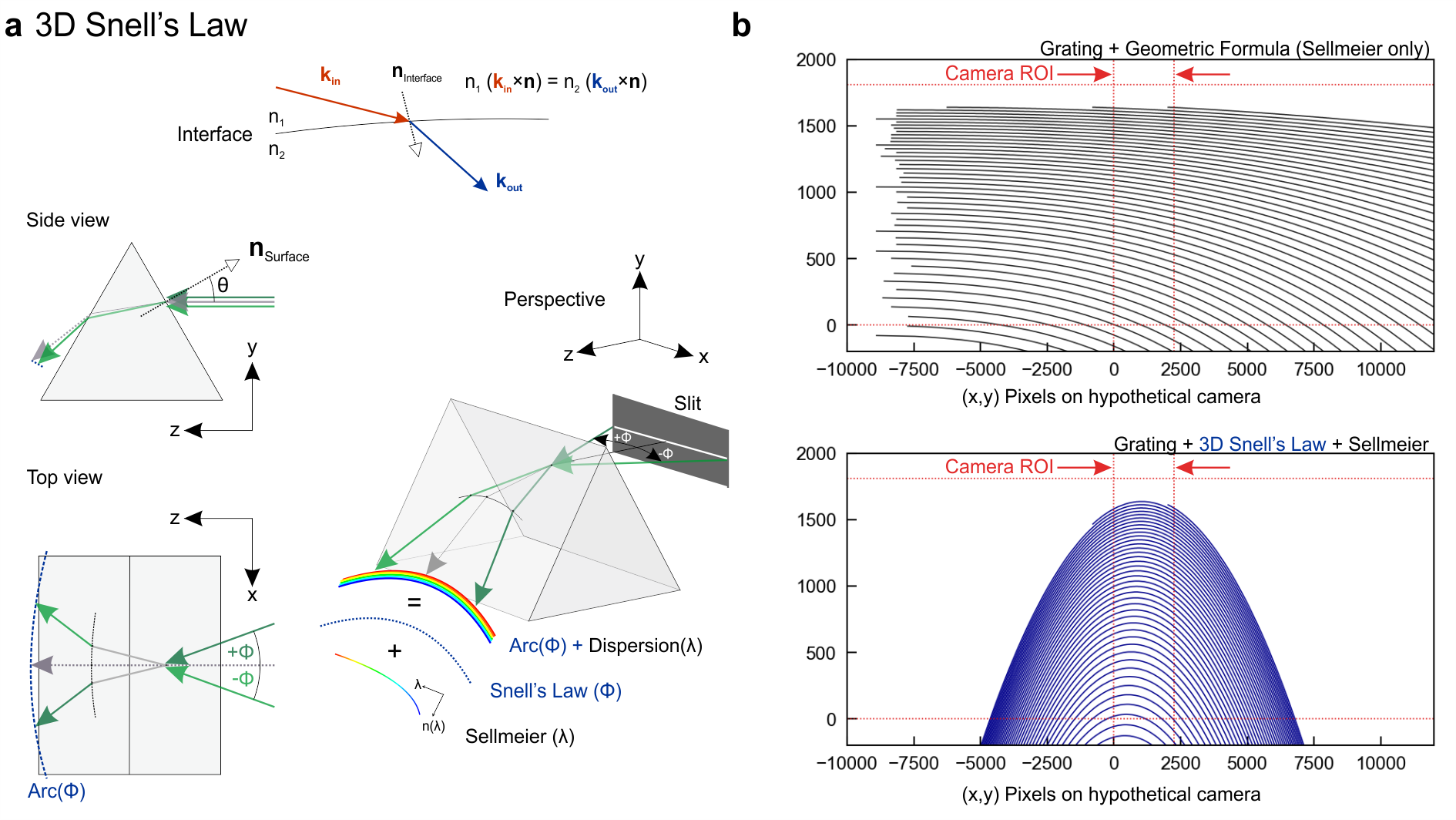}
	\caption{Snell’s Law in three dimensions. \textbf{a}. \textit{Upper panel}: Vector-form of Snell’s law describing the refraction of light beams traversing the interface between two media $n_1$ and $n_2$. \textit{Lower panel}: Dispersion of white light in prisms. Light from a line-source entering a prism will be dispersed and warped into a bow due to geometrical effects. \textbf{b}. Simulated cross-dispersed \'{E}chelle spec\-tro\-grams using the geometric dispersion equation for prisms (upper panel) and simulated spectra using the three-dimensional equation for Snell’s law (lower panel).}
	\label{fig:figure2}
\end{figure*}
\subsection{Snell's law in vector form}
\noindent In the previous example, we considered the prism's entry facet normal vector to be in the same plane as the in- and outgoing light.
In general, this assumption is not applicable and the geometric prism equation (\ref{eq:prismgeo}) breaks down.
How do we want to tackle this issue?\newline
Our approach is applying Snell’s law in three dimensions to the entry and exit facets of the prism using:
\begin{equation}
	n_1\left( \mathbf{k}_{ in }\times \mathbf{n}\right)=n_2\left( \mathbf{k}_{ out }\times \mathbf{n} \right),
\end{equation}
where $n_1$, $n_2$ are the refractive indices of two adjacent media, $\mathbf{k}_{in}$, $\mathbf{k}_{out}$ are the normalized incoming and outgoing beam direction vectors while $\mathbf{n}$ is the normal vector of the entry surface.
Using vector algebra identities, $\mathbf{k}_{out}$ can be represented as a function of $\mathbf{n}$ and $\mathbf{k}_{in}$ (see supp\-le\-men\-ta\-ry material section~\ref{supp:snellslawopus}):

\begin{widetext}
\begin{eqnarray}
	\mathbf{k}_{out}\left(\mathbf{n},\mathbf{k}_{in}\right)=
	\left(\sqrt{1-\left(\frac{n_1}{n_2}\right)^2\left[1-\left( \mathbf{n} \cdot \mathbf{k}_{in}\right)^2\right]}-\frac{n_1}{n_2} \left(\mathbf{n}\cdot\mathbf{k}_{in}\right)\right)\cdot \mathbf{n} +\frac{n_1}{n_2} \mathbf{k}_{in}
	\label{eq:prismraytrace}
\end{eqnarray}
\end{widetext}
\noindent This equation does not contain tedious trigonometric functions but rather consists of scalar products that can be computed fairly fast \cite{bib:Miks2012Snell,bib:frobenius2018} which is why most ray tracing applications \cite{bib:Whitted1980,bib:Glassner1989} rely on this formula as it turned out during our literature research.

We summarised our findings in Figure \ref{fig:figure2}.
The upper panel in \textbf{a} depicts a light beam crossing an interface.
The incoming ray $\mathbf{k}_{in}$ (red arrow) traverses the interface between two media with refractive indices $n_1$ and $n_2$ with $n_2 > n_1$, respectively.
It is refracted towards the surface normal vector $\mathbf{n}$ according to Snell's law.
The resulting light ray is represented by the blue arrow $\mathbf{k}_{out}$.
Without loss of generality, we apply this equation to an equilateral prism as shown in the lower panel of \figref{fig:figure2}{a}.

\noindent Here, we choose a point on the prism and model a small horizontal slit by two beams.
These beams emerge from the edges of the slit at angles $\pm\phi$ relative to the z-axis and an entry point on the prism.
Equation (\ref{eq:prismraytrace}) dictates that the incoming beams will be refracted onto an arc (dashed black inner arc, see supplementary information).
This indeed also applies when both beams leave the prism (blue dashed  arc).
Since different colours are dispersed into slightly different angles, the resulting arc will be dispersed along the z- and negative y-direction in our laboratory frame as a result of a wavelength dependent refractive index (Sell\-meier e\-qua\-tion).
This behaviour cannot be completely modelled via the frequently used trigonometric equation (\ref{eq:prismgeo}).
To visually understand this better, we prepared an interactive online example using the tool on \emph{Math3D.org} \cite{bib:Wierzbowski2022}. A Python implementation and a simple calculation can be found in the appendix.\newline
\indent Dispersing and warping the incident light into an arc, the prism has a strong effect on spec\-tro\-grams taken with \'{E}chelle spectrometers in a Grating-Prism configuration.
\figref{fig:figure2}{b} contrasts computed spectra using the trigonometric prism equation (\ref{eq:prismgeo}) and the vectorial equation (\ref{eq:prismraytrace}).
The upper panel is a simulated cross-dispersed spectrogram (black) which was obtained using the geometrical equation (\ref{eq:prismgeo}) whilst the lower panel depicts a spectrogram simulated with the vectorial approach (blue).
We compressed the x-direction of the detector to visualise the differences also outside of a typical camera region of interest (ROI).
In the upper panel of \figref{fig:figure2}{b}, we can clearly see all spectral orders having a monotonically decreasing behaviour for ever larger x-values.
In our calculations, shorter wavelengths of a respective order appear at larger x-pixel values.
We can observe the direct influence of the Sellmeier equation on the dispersion behaviour, since equation (\ref{eq:prismgeo}) only accounts for $\theta$, the entry angle in the yz-plane, and the wavelength dependent refractive index.\newline
The spectrogram strongly changes in the second case as Snell's Law in it's vector form accounts for both, $\theta$ and $\phi$.
The latter being the entry angle in the xz-plane (see Fig.~\ref{fig:figure2}\textbf{a}, lower left panel).
This method correctly describes and predicts the shape and positions of spectral orders on a detector.
Please note, that the calculated spectra depend on geometrical parameters such as grating position, prism angles, detector rotation, etc.
Thus, fitting spectra or single wavelengths in a measured spectrum enables the determination of the actual angles and positions of the optical components.\newline
ThAr cold-cathode-lamps with distinct single peaks are a great candidate for this.
These lamps are typically used to calibrate \'{E}chelle spectrograms \cite{bib:Schanne2018} and are within the standard toolset of professional and enthusiast astronomers.
Our approach has the potential to save time during calibration and enables algorithms for auto-calibrations or tracing thermal drifts in longer measurement campaigns.
\section{Comparison and Outlook}
\begin{figure*}[ht]
	\centering
	\includegraphics[width=1\textwidth,keepaspectratio]{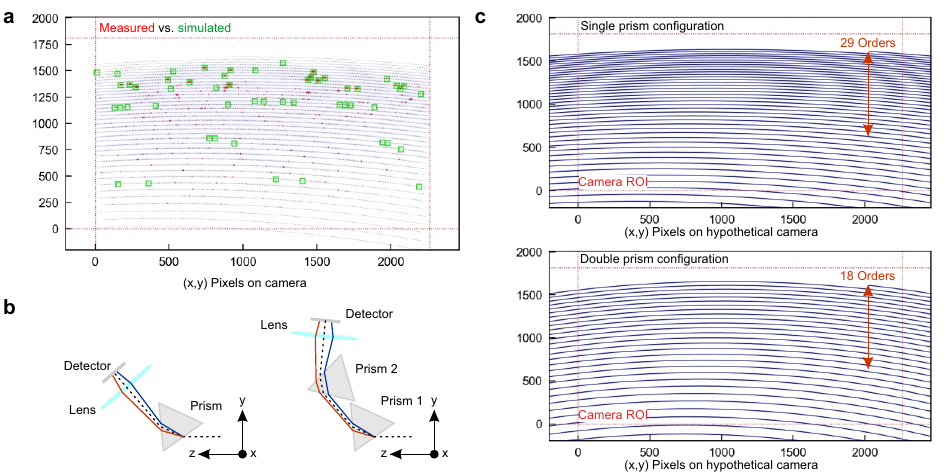}
	\caption{Comparison of simulation and measurement. \textbf{a}. Overlay of a measured flat field spectrum of a tungsten lamp on a FLECHAS spectrometer (light blue) and a simulated \'{E}chelle spectrum (black, dashed lines). The red dots are measured ThAr peaks. The green boxes are simulated ThAr peak positions for selected Argon and Thorium lines from \emph{nist.gov}. The overlap is very good over all visible orders. This good agreement is achieved through proper consideration of the spectrometer geometry in the simulation. \textbf{b}. Single prism and double prism cross-disperser arrangement. The double-prism configuration is used e.g. in white pupil spectrometers to achieve good separation of the orders and folding the beam path. \textbf{c}. Simulation of a spectrogram with a single prism cross-disperser (upper panel) and double-prism configuration (lower panel). The red arrow of constant length helps to show how the second prism introduces more dispersion and a stronger bending behaviour of the spectral orders towards lower x pixel values.}
	\label{fig:figure3}
\end{figure*}

\noindent In \figref{fig:figure3}{a}, we compare our findings using the vectorial representation of Snell’s law with real \'{E}chelle spectra taken with a FLECHAS spectrograph, designed by Carlos Guirao and colleagues \cite{bib:Flechas2010}\cite{bib:Mugrauer2014}.
In our case, the spectrometer is equipped with a $d=79$ lines/mm grating, a $\SI{60}{\milli\meter}$ Schott glass N-F2 prism, an $f = \SI{200}{\milli\meter}$ objective lens as well as a $4524\times3624$ pixel camera with a pixel size of $\SI{6}{\micro\meter}$.
The spectra and calculations were carried out using a full $2\times2$ pixel binning.
To perform the calculations properly, we extracted the spectrometer geometry from available CAD files \cite{bib:Flechas2010}.
For contrast reasons, we plotted a measured spec\-tro\-gram of a table-top tungsten lamp (flat field spectrum) in light blue and overlaid a calculated spectrogram with dashed black lines using the vectorial approach.
The red dots in the image correspond to measured ThAr peaks on the same spectrometer.
At the same time, the green boxes mark simulated spectral positions for selected Ar and Th catalogue lines from NIST \cite{bib:NIST2023}.
Here, the calculated orders range from $\left|m\right| = 19\ \text{to}\ 57$.
The overlap between these spec\-tro\-grams is very good due to proper spec\-tro\-me\-ter alignment and slightly adjusted simulation parameters (detector rotation, magnification, etc.) based on the spectrometer geometry. An enlarged version of \figref{fig:figure3}{a} can be found in the supplementary information (\figref{fig:suppfigure5}{}).\newline
We will make our simulation package \href{https://github.com/jw-echelle/QtYETI}{QtYETI} \cite{bib:qtyeti2023} available on github which includes the vectorial approach.\newline
This satisfying result of course leads to the question, if the algorithm is applicable to cross-dispersers with a serial arrangement of dispersion prisms.
\figref{fig:figure3}{b} shows two possible scenarios.
The left sketch shows a single prism configuration that can be found in FLECHAS and other \'{E}chelle spectrometers.
Light that has been dispersed in the xz-plane by a grating hits the prism facet and is then further dispersed in the yz-plane and eventually focused on a detector.
The resulting spectrogram is shown in \figref{fig:figure3}{c} (upper panel). Within an arbitrary, yet, fixed interval (red arrow) we find 29 orders.
The spectrogram changes significantly, if we now introduce another prism (Fig. \textbf{\ref{fig:figure3}b}, lower right panel, N-F2 glass) and change the camera angle such that the beginning of the 20th order starts at the same pixel (see supplementary information).
Within the same fixed interval as before, we clearly see a wider dispersion resulting in only 18 orders.
This is of course to be expected as another prism further disperses the incoming light in the yz-plane.
This double prism configuration can be found in white-pupil spectrometers to further fold the beam path while increasing dispersion.
Further details can be found in \cite{bib:Sablowski2018}.\vfill

\section{Summary}
\noindent In this work, we have shown how two-di\-men\-sion\-al \'{E}chelle spec\-tro\-grams emerge from using a gra\-ting-prism cross-dis\-per\-sion configuration.
We used Snell’s law in it's vector from as base for our calculations and correctly predicted the shape and position of grating diffraction orders for grating-prism \'{E}chelle spectrographs.
We have found significant differences between using the common trigonometric prism dispersion equation and a full vectorial representation of Snell's law.
We continued to show a quantitatively good overlap between calculated and measured spectra for a FLECHAS spectrograph.
In the last part, we applied our simulation code to a double-prism configuration with reasonable results, which could easily be verified in a laboratory setup. 

In the next steps we will continue our work to accomplish a robust way to auto-calibrate a spectrometer and \'{E}chelle spectra using a ThAr-Lamp (thorium argon).
Our algorithms will be implemented in the user-friendly cross-platform open-source data reduction software \href{https://github.com/jw-echelle/QtYETI}{QtYETI} written in Python.\newline
The authors want to thank Lothar Schanne, Klaus Vollmann, Ulrich Waldschlaeger, Herbert P\"{u}hringer and Daniel Sablowski for valuable inputs and discussions as well as corrections from Annette F\"{a}lschle, Lukas Hanschke \& Friedrich Sbresny.
\bibliographystyle{apsrev4-2}
\bibliography{OnPrismCrossDispersers}

\begin{thebibliography}{19}%
\makeatletter
\providecommand \@ifxundefined [1]{%
 \@ifx{#1\undefined}
}%
\providecommand \@ifnum [1]{%
 \ifnum #1\expandafter \@firstoftwo
 \else \expandafter \@secondoftwo
 \fi
}%
\providecommand \@ifx [1]{%
 \ifx #1\expandafter \@firstoftwo
 \else \expandafter \@secondoftwo
 \fi
}%
\providecommand \natexlab [1]{#1}%
\providecommand \enquote  [1]{``#1''}%
\providecommand \bibnamefont  [1]{#1}%
\providecommand \bibfnamefont [1]{#1}%
\providecommand \citenamefont [1]{#1}%
\providecommand \href@noop [0]{\@secondoftwo}%
\providecommand \href [0]{\begingroup \@sanitize@url \@href}%
\providecommand \@href[1]{\@@startlink{#1}\@@href}%
\providecommand \@@href[1]{\endgroup#1\@@endlink}%
\providecommand \@sanitize@url [0]{\catcode `\\12\catcode `\$12\catcode
  `\&12\catcode `\#12\catcode `\^12\catcode `\_12\catcode `\%12\relax}%
\providecommand \@@startlink[1]{}%
\providecommand \@@endlink[0]{}%
\providecommand \url  [0]{\begingroup\@sanitize@url \@url }%
\providecommand \@url [1]{\endgroup\@href {#1}{\urlprefix }}%
\providecommand \urlprefix  [0]{URL }%
\providecommand \Eprint [0]{\href }%
\providecommand \doibase [0]{https://doi.org/}%
\providecommand \selectlanguage [0]{\@gobble}%
\providecommand \bibinfo  [0]{\@secondoftwo}%
\providecommand \bibfield  [0]{\@secondoftwo}%
\providecommand \translation [1]{[#1]}%
\providecommand \BibitemOpen [0]{}%
\providecommand \bibitemStop [0]{}%
\providecommand \bibitemNoStop [0]{.\EOS\space}%
\providecommand \EOS [0]{\spacefactor3000\relax}%
\providecommand \BibitemShut  [1]{\csname bibitem#1\endcsname}%
\let\auto@bib@innerbib\@empty
\bibitem [{\citenamefont {{Schroeder}}\ and\ \citenamefont
  {{Anderson}}(1971)}]{bib:Schroeder1971}%
  \BibitemOpen
  \bibfield  {author} {\bibinfo {author} {\bibfnamefont {D.~J.}\ \bibnamefont
  {{Schroeder}}}\ and\ \bibinfo {author} {\bibfnamefont {C.~M.}\ \bibnamefont
  {{Anderson}}},\ }\href {https://doi.org/10.1086/129150} {\bibfield  {journal}
  {\bibinfo  {journal} {PASP}\ }\textbf {\bibinfo {volume} {83}},\ \bibinfo
  {pages} {438} (\bibinfo {year} {1971})}\BibitemShut {NoStop}%
\bibitem [{\citenamefont {{Vogt}}(1987)}]{bib:Hamilton1987}%
  \BibitemOpen
  \bibfield  {author} {\bibinfo {author} {\bibfnamefont {S.~S.}\ \bibnamefont
  {{Vogt}}},\ }\href {https://doi.org/10.1086/132107} {\bibfield  {journal}
  {\bibinfo  {journal} {PASP}\ }\textbf {\bibinfo {volume} {99}},\ \bibinfo
  {pages} {1214} (\bibinfo {year} {1987})}\BibitemShut {NoStop}%
\bibitem [{\citenamefont {{Appenzeller}}\ \emph {et~al.}(1988)\citenamefont
  {{Appenzeller}}, \citenamefont {{Krautter}}, \citenamefont {{Mandel}},\ and\
  \citenamefont {{Oestreicher}}}]{bib:Orfeus1988}%
  \BibitemOpen
  \bibfield  {author} {\bibinfo {author} {\bibfnamefont {I.}~\bibnamefont
  {{Appenzeller}}}, \bibinfo {author} {\bibfnamefont {J.}~\bibnamefont
  {{Krautter}}}, \bibinfo {author} {\bibfnamefont {H.}~\bibnamefont
  {{Mandel}}},\ and\ \bibinfo {author} {\bibfnamefont {R.}~\bibnamefont
  {{Oestreicher}}},\ }in\ \href@noop {} {\emph {\bibinfo {booktitle} {ESA
  Special Publication}}},\ \bibinfo {series} {ESA Special Publication},
  Vol.~\bibinfo {volume} {2},\ \bibinfo {editor} {edited by\ \bibinfo {editor}
  {\bibfnamefont {N.}~\bibnamefont {{Longdon}}}\ and\ \bibinfo {editor}
  {\bibfnamefont {E.~J.}\ \bibnamefont {{Rolfe}}}}\ (\bibinfo {year} {1988})\
  pp.\ \bibinfo {pages} {337--340}\BibitemShut {NoStop}%
\bibitem [{\citenamefont {{Strassmeier}}\ \emph {et~al.}(2018)\citenamefont
  {{Strassmeier}}, \citenamefont {{Ilyin}},\ and\ \citenamefont
  {{Steffen}}}]{bib:Pepsi2018}%
  \BibitemOpen
  \bibfield  {author} {\bibinfo {author} {\bibfnamefont {K.~G.}\ \bibnamefont
  {{Strassmeier}}}, \bibinfo {author} {\bibfnamefont {I.}~\bibnamefont
  {{Ilyin}}},\ and\ \bibinfo {author} {\bibfnamefont {M.}~\bibnamefont
  {{Steffen}}},\ }\href {https://doi.org/10.1051/0004-6361/201731631}
  {\bibfield  {journal} {\bibinfo  {journal} {AAP}\ }\textbf {\bibinfo {volume}
  {612}},\ \bibinfo {eid} {A44} (\bibinfo {year} {2018})},\ \Eprint
  {https://arxiv.org/abs/1712.06960} {arXiv:1712.06960 [astro-ph.SR]}
  \BibitemShut {NoStop}%
\bibitem [{\citenamefont {Palmer}(2014)}]{bib:Palmer2014}%
  \BibitemOpen
  \bibfield  {author} {\bibinfo {author} {\bibfnamefont {C.}~\bibnamefont
  {Palmer}},\ }\href@noop {} {\emph {\bibinfo {title} {Diffraction Grating
  Handbook (7th edition)}}}\ (\bibinfo {year} {2014})\BibitemShut {NoStop}%
\bibitem [{\citenamefont {{Schanne}}\ and\ \citenamefont
  {{Sablowski}}(2018)}]{bib:Schanne2018}%
  \BibitemOpen
  \bibfield  {author} {\bibinfo {author} {\bibfnamefont {L.}~\bibnamefont
  {{Schanne}}}\ and\ \bibinfo {author} {\bibfnamefont {D.}~\bibnamefont
  {{Sablowski}}},\ }\href@noop {} {\emph {\bibinfo {title}
  {{A}strophysikalische {I}nstrumentierung und {M}esstechnik f\"{u}r die
  {S}pektroskopie: {T}heorie, {P}raxis, {T}echnik und {B}eobachtung}}},\
  \bibinfo {edition} {1st}\ ed.\ (\bibinfo  {publisher} {Schanne, Lothar F},\
  \bibinfo {address} {Mannheim, Germany},\ \bibinfo {year} {2018})\BibitemShut
  {NoStop}%
\bibitem [{\citenamefont {Griffiths}(2013)}]{bib:griffiths2013introduction}%
  \BibitemOpen
  \bibfield  {author} {\bibinfo {author} {\bibfnamefont {D.~J.}\ \bibnamefont
  {Griffiths}},\ }\href@noop {} {\emph {\bibinfo {title} {Introduction to
  {E}lectrodynamics}}}\ (\bibinfo  {publisher} {Pearson},\ \bibinfo {year}
  {2013})\BibitemShut {NoStop}%
\bibitem [{\citenamefont {von Sellmeier}(1871)}]{bib:Sellmeier1871}%
  \BibitemOpen
  \bibfield  {author} {\bibinfo {author} {\bibfnamefont {W.}~\bibnamefont {von
  Sellmeier}},\ }\href
  {https://doi.org/https://doi.org/10.1002/andp.18712190612} {\bibfield
  {journal} {\bibinfo  {journal} {Annalen der Physik}\ }\textbf {\bibinfo
  {volume} {219}},\ \bibinfo {pages} {272} (\bibinfo {year}
  {1871})}\BibitemShut {NoStop}%
\bibitem [{\citenamefont {Miks}\ and\ \citenamefont
  {Novak}(2012)}]{bib:Miks2012Snell}%
  \BibitemOpen
  \bibfield  {author} {\bibinfo {author} {\bibfnamefont {A.}~\bibnamefont
  {Miks}}\ and\ \bibinfo {author} {\bibfnamefont {P.}~\bibnamefont {Novak}},\
  }\href {https://doi.org/10.1364/JOSAA.29.001356} {\bibfield  {journal}
  {\bibinfo  {journal} {JOSA A}\ }\textbf {\bibinfo {volume} {29}},\ \bibinfo
  {pages} {1356} (\bibinfo {year} {2012})}\BibitemShut {NoStop}%
\bibitem [{\citenamefont {{Frobenius}}\ and\ \citenamefont {{Stack
  Exchange}}(2018)}]{bib:frobenius2018}%
  \BibitemOpen
  \bibfield  {author} {\bibinfo {author} {\bibnamefont {{Frobenius}}}\ and\
  \bibinfo {author} {\bibnamefont {{Stack Exchange}}},\ }\href@noop {}
  {\bibinfo {title} {Snell's law in vector form}},\ \bibinfo {howpublished}
  {\url{https://physics.stackexchange.com/questions/435512/snells-law-in-vector-form}}
  (\bibinfo {year} {2018}),\ \bibinfo {note} {accessed: 2023-5-3}\BibitemShut
  {NoStop}%
\bibitem [{\citenamefont {Whitted}(1980)}]{bib:Whitted1980}%
  \BibitemOpen
  \bibfield  {author} {\bibinfo {author} {\bibfnamefont {T.}~\bibnamefont
  {Whitted}},\ }\href {https://doi.org/10.1145/358876.358882} {\bibfield
  {journal} {\bibinfo  {journal} {Commun. ACM}\ }\textbf {\bibinfo {volume}
  {23}},\ \bibinfo {pages} {343–349} (\bibinfo {year} {1980})}\BibitemShut
  {NoStop}%
\bibitem [{\citenamefont {Glassner}(1989)}]{bib:Glassner1989}%
  \BibitemOpen
  \bibinfo {editor} {\bibfnamefont {A.~S.}\ \bibnamefont {Glassner}},\ ed.,\
  \href@noop {} {\emph {\bibinfo {title} {An introduction to ray tracing}}}\
  (\bibinfo  {publisher} {Morgan Kaufmann},\ \bibinfo {address} {Oxford,
  England},\ \bibinfo {year} {1989})\BibitemShut {NoStop}%
\bibitem [{\citenamefont {{Wierzbowski}}(2022)}]{bib:Wierzbowski2022}%
  \BibitemOpen
  \bibfield  {author} {\bibinfo {author} {\bibfnamefont {J.}~\bibnamefont
  {{Wierzbowski}}},\ }\href@noop {} {\bibinfo {title} {{P}rism refraction via
  {S}nell's {L}aw in 3{D}}},\ \bibinfo {howpublished}
  {\url{https://www.math3d.org/b4TzRiJ1M}} (\bibinfo {year} {2022})\BibitemShut
  {NoStop}%
\bibitem [{\citenamefont {{Guirao}}(2010)}]{bib:Flechas2010}%
  \BibitemOpen
  \bibfield  {author} {\bibinfo {author} {\bibfnamefont {C.}~\bibnamefont
  {{Guirao}}},\ }\href@noop {} {\bibinfo {title} {Integrating {FLECHAS}, a
  breadboard {E}chelle spectrograph}},\ \bibinfo {howpublished}
  {\url{https://spectroscopy.wordpress.com/2010/05/14/integrating-an-optical-bench-echelle-spectrograph/}}
  (\bibinfo {year} {2010})\BibitemShut {NoStop}%
\bibitem [{\citenamefont {Mugrauer}\ \emph {et~al.}()\citenamefont {Mugrauer},
  \citenamefont {Avila},\ and\ \citenamefont {Guirao}}]{bib:Mugrauer2014}%
  \BibitemOpen
  \bibfield  {author} {\bibinfo {author} {\bibfnamefont {M.}~\bibnamefont
  {Mugrauer}}, \bibinfo {author} {\bibfnamefont {G.}~\bibnamefont {Avila}},\
  and\ \bibinfo {author} {\bibfnamefont {C.}~\bibnamefont {Guirao}},\
  }\href@noop {} {\bibfield  {journal} {\bibinfo  {journal} {Astronomische
  Nachrichten}\ }\textbf {\bibinfo {volume} {335}},\ \bibinfo {pages}
  {417}}\BibitemShut {NoStop}%
\bibitem [{\citenamefont {{NIST - National Institute of Standards and
  Technology}}(2023)}]{bib:NIST2023}%
  \BibitemOpen
  \bibfield  {author} {\bibinfo {author} {\bibnamefont {{NIST - National
  Institute of Standards and Technology}}},\ }\href
  {https://physics.nist.gov/PhysRefData/Handbook/element_name.htm} {\bibinfo
  {title} {Basic atomic spectroscopic data.}} (\bibinfo {year}
  {2023})\BibitemShut {NoStop}%
\bibitem [{\citenamefont {Wierzbowski}(2023)}]{bib:qtyeti2023}%
  \BibitemOpen
  \bibfield  {author} {\bibinfo {author} {\bibfnamefont {J.}~\bibnamefont
  {Wierzbowski}},\ }\href {https://github.com/jw-echelle/QtYETI} {\bibinfo
  {title} {{QtYETI - Yeti's Extra-Terrestrial Investigations}}} (\bibinfo
  {year} {2023})\BibitemShut {NoStop}%
\bibitem [{\citenamefont {Sablowski}\ \emph {et~al.}(2018)\citenamefont
  {Sablowski}, \citenamefont {Woche}, \citenamefont {Weber}, \citenamefont
  {J{\"a}rvinen},\ and\ \citenamefont {Strassmeier}}]{bib:Sablowski2018}%
  \BibitemOpen
  \bibfield  {author} {\bibinfo {author} {\bibfnamefont {D.~P.}\ \bibnamefont
  {Sablowski}}, \bibinfo {author} {\bibfnamefont {M.}~\bibnamefont {Woche}},
  \bibinfo {author} {\bibfnamefont {M.}~\bibnamefont {Weber}}, \bibinfo
  {author} {\bibfnamefont {A.}~\bibnamefont {J{\"a}rvinen}},\ and\ \bibinfo
  {author} {\bibfnamefont {K.~G.}\ \bibnamefont {Strassmeier}},\ }in\ \href
  {https://doi.org/10.1117/12.2312476} {\emph {\bibinfo {booktitle} {Advances
  in Optical and Mechanical Technologies for Telescopes and Instrumentation
  III}}},\ Vol.\ \bibinfo {volume} {10706},\ \bibinfo {editor} {edited by\
  \bibinfo {editor} {\bibfnamefont {R.}~\bibnamefont {Navarro}}\ and\ \bibinfo
  {editor} {\bibfnamefont {R.}~\bibnamefont {Geyl}}},\ \bibinfo {organization}
  {International Society for Optics and Photonics}\ (\bibinfo  {publisher}
  {SPIE},\ \bibinfo {year} {2018})\ p.\ \bibinfo {pages} {107066F}\BibitemShut
  {NoStop}%
\bibitem [{\citenamefont {{Schott AG}}(2023)}]{bib:Schott2023}%
  \BibitemOpen
  \bibfield  {author} {\bibinfo {author} {\bibnamefont {{Schott AG}}},\ }\href
  {https://www.schott.com} {\bibinfo {title} {Refractive indices of prisms.}}
  (\bibinfo {year} {2023})\BibitemShut {NoStop}%
\end{thebibliography}%
\onecolumngrid
\newpage

\section{Supplementary Material}
\subsection{Sellmeier's Equation}

\noindent The Sellmeier equation describes the refractive index of glasses within visible wavelength range and slightly beyond \cite{bib:Sellmeier1871}.
$$n\left(\lambda\right)=\sqrt{1+\sum_{i=1,..,n}\frac{B_i\ \lambda^2}{\lambda^2-C_i}}.$$

\noindent In this work, we used the below coefficients to successfully model our \'{E}chelle spectra which are taken from \cite{bib:Schott2023}

\begin{center}
\begin{tabular}{l | l | l}
	Coefficients & $B_i$ & $C_i$ \\
	\toprule
	N-F2
	& B1 = 1.34533359 & C1 = 0.00997743871\\
	& B2 = 0.209073176 & C2 = 0.0470450767 \\
	& B3 = 0.93757162 & C3 = 111.8867640\\
	\midrule
	N-NF2
	& B1 = 1.39757037 & C1 = 0.00995906143\\
	& B2 = 0.159201403 & C2 = 0.0546931752\\
	& B3 = 1.2686543 & C3 = 119.248346
\end{tabular}
\end{center}

\subsection{Grating equation}
\noindent The derivation and complete discussion of the grating equation can be found in the MKS Grating Handbook which can be found on https://www.newport.com
\begin{figure*}[b]
	\centering
	\includegraphics[keepaspectratio]{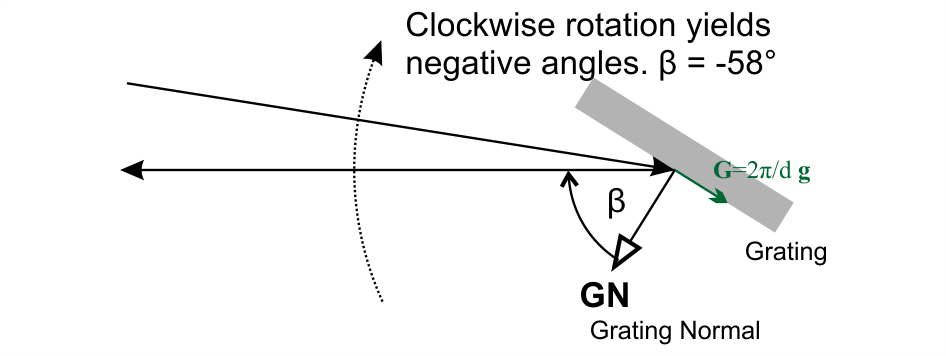}
	\caption{Grating angle definition.}
	\label{fig:suppfigure1}
\end{figure*}
$$m\cdot\lambda=d( \sin\ \alpha\ +\ \sin\ \beta)\cos\ \gamma$$
We use the following convention: every angle is measured from the grating normal GN towards any beam of light.
If the rotation is clockwise or counter-clockwise, the corresponding angle is negative or positive, respectively.
$\alpha$ is the angle of incidence relative to the grating normal.
$\beta$ is the outgoing angle of the diffracted light for an order $m$ and $m$ is the grating constant.\newline
Example: A 79 grooves/$\unit{\micro\meter}$ grating exhibits a grating constant $d = 1/79\unit{\milli\meter}$ = $~\SI{12}{\micro\meter}$/groove.
The angle $\gamma$, which we treated as $\gamma = \deg{0}$ for simplicity, is the out-of-plane angle of light with respect to the plane of incidence of a grating (if the grating vector $G = (2\pi/d) \cdot \mathbf{g}$ lies perfectly parallel to the xz-plane).
$\mathbf{G}$ is the reciprocal grating vector.
For constant observation angles, the right-hand side of grating equation $m\cdot\lambda=d(\sin\ \alpha + \sin\ \beta)\cos\ \gamma$
becomes constant and can be solved for multiple combinations of m and $\lambda$. Suppose $\alpha = \SI{-67}{\degree}$,  $\beta = \SI{-58}{\degree}$, $\gamma = \SI{0}{\degree}$, $d = 79$ grooves/$\unit{\micro\meter}$, then $m\cdot\lambda = \SI{-22.30}{\micro\meter}$. If $m = -20, -21, -22,...$ and $\lambda = \SI{1,115}{\micro\meter}, \SI{1,062}{\micro\meter}, \SI{1,014}{\micro\meter},...$
\vfill\eject

\subsection{Derivation of Snell’s law into a scalar product equation}\label{supp:snellslawopus}

\begin{figure*}[!]
	\centering
	\includegraphics[width=100mm,keepaspectratio]{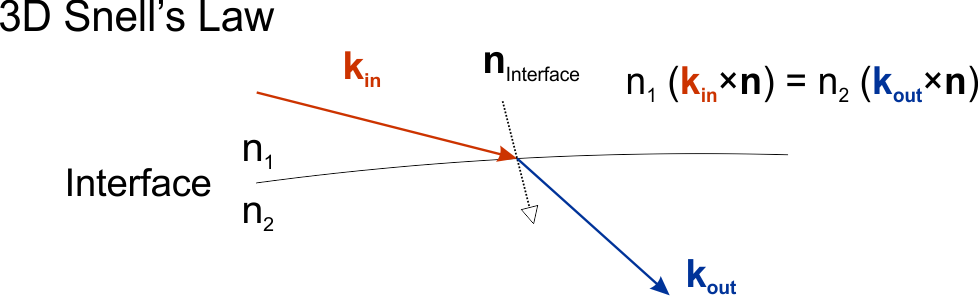}
	\caption{General application of Snell's law.}
	\label{fig:suppfigure2}
\end{figure*}

\noindent Snell’s law can be expressed as:
\begin{align}
	n_1\left( \mathbf{n} \times \mathbf{k}_{in}\right)=n_2\left( \mathbf{n} \times \mathbf{k}_{out}\right)\nonumber
\end{align}
\noindent Please note, we will use the general definition $\mathbf{k} = \mathbf{k}^{ip} + \mathbf{k}^{op}$, where $\mathbf{k}^{ip} = \mathbf{k}- \left(\mathbf{n}\cdot\mathbf{k}\right) \mathbf{n}$ and $\mathbf{k}^{op} = \left(\mathbf{n}\cdot\mathbf{k}\right) \mathbf{n}$ are the in-plane and out-of-plane components of a generic vector $\mathbf{k}$ with respect to a surface normal vector $\mathbf{n}$.\newline

\noindent With the relation $\mathbf{a}\times \mathbf{b}=\left| \mathbf{a}\right|\left|\mathbf{b}\right|\ sin\ \theta$, we obtain the known relation $n_1 \sin \theta_1 = n_2 \sin \theta_2$, as $\mathbf{k}$ and $\mathbf{n}$ are of unit length.  By applying $(\mathbf{n} \times)$ to both sides of the upper equation and by means of the vector identity $\mathbf{A}\times(\mathbf{B}\times\mathbf{C}) = (\mathbf{A}\cdot\mathbf{C})\times\mathbf{B}\ –\ (\mathbf{A}\cdot\mathbf{B})\times\mathbf{C}$, we can transform Snell’s law into:
\begin{align}
	-\frac{n_1}{n_2}\left[ \mathbf{k}_{in}-\left( \mathbf{n} \cdot \mathbf{k}_{in}\right) \mathbf{n} \right]= \mathbf{k}_{out}-\left( \mathbf{n} \cdot \mathbf{k}_{out}\right) \mathbf{n} =: \mathbf{k}_{out}^{ip}\nonumber
\end{align}
Both sides of the equation correspond to in-plane components of $\mathbf{k}_{in}$ and $\mathbf{k}_{out}$, respectively.
This results from $\mathbf{k}^{op} = \left(\mathbf{n}\cdot\mathbf{k}\right) \mathbf{n}$ being the out-of-plane component of a vector $\mathbf{k}$.
We can now express the in-plane component of $\mathbf{k}_{out}^{ip}$ in terms of $\mathbf{k}_{in}$ and the refractive indices.
The next step is calculating the out-of-plane component of $\mathbf{k} _{out}^{op}$ as a function of $\mathbf{k}_{in}$.\newline

\noindent The squared length of $\mathbf{k}_{out}$ can be written as:
\begin{align}
	k_{out}^2 &= \mathbf{k}_{out}^{op}\cdot \mathbf{k}_{out}^{op} + \mathbf{k}_{out}^{ip}\cdot\mathbf{k}_{out}^{ip}\nonumber\\
	&=\mathbf{k}_{out}^{op}\cdot \mathbf{k}_{out}^{op} + (\frac{n_1}{n_2}\left[\left( \mathbf{n} \cdot \mathbf{k}_{in}\right) \mathbf{n} - \mathbf{k}_{in}\right])(\frac{n_1}{n_2}\left[\left( \mathbf{n} \cdot \mathbf{k}_{in}\right) \mathbf{n} - \mathbf{k}_{in}\right])\nonumber
\end{align}
\noindent with $\mathbf{k}^{ip} \cdot \mathbf{k}^{op} = 0$, since  and thus both vectors are defined as orthogonal to one another.\newline
\noindent After multiplicative expansion and subsequent simplification of the upper expression, we obtain:
$$k_{out}^2=\mathbf{k}_{out}^{op}\cdot \mathbf{k}_{out}^{op} + \left(\frac{n_1}{n_2}\right)^2\left[ \mathbf{k}_{in}\cdot \mathbf{k}_{in}-\left( \mathbf{n} \cdot \mathbf{k}_{in}\right)^2\right]$$
Let $| \mathbf{k}_{in}|\ =\ | \mathbf{k}_{out}|=1$, such that:
\begin{align}
	k_{out}^{op}=\sqrt{1-\left(\frac{n_1}{n_2}\right)^2\left[1-\left( \mathbf{n} \cdot \mathbf{k}_{in}\right)^2\right]}\nonumber.
\end{align}
\noindent Using this result, we can fully express $\mathbf{k}_{out}$ as a function of the surface normal vector $\mathbf{n}$ and the incident light beam direction vector $\mathbf{k}_{in}$:
$$\mathbf{k}_{out}\left(\mathbf{n},\mathbf{k}_{in}\right)=\left(\sqrt{1-\left(\frac{n_1}{n_2}\right)^2\left[1-\left( \mathbf{n} \cdot \mathbf{k}_{in}\right)^2\right]}-\left(\frac{n_1}{n_2}\right) \left(\mathbf{n} \cdot \mathbf{k}_{in}\right)\right)\cdot \mathbf{n} +\left(\frac{n_1}{n_2}\right) \mathbf{k}_{in}$$
This expression solely depends on scalar products enabling fast computation instead of lengthy trigonometric relations. During literature research, we found that this formula is/was used in efficient raytracing applications.
We have created an interactive math3d.org project \cite{bib:Wierzbowski2022} to help the reader visualizing all the relationships.
During literature research, we found that raytracing applications use this kind of formula, which is vastly unknown in spectroscopy.
Further reading can be found in \cite{bib:Whitted1980} \cite{bib:Glassner1989} \cite{bib:Miks2012Snell} \cite{bib:frobenius2018}.
\vfill\eject\newpage

\subsection{Snell's law - example calculation}
\noindent The arcing effect seen in our prism example in the main text only arises from an oblique incidence on an optical surface.
We will now calculate a simple example of a light beam travelling in vacuum $(n_1 = 1)$ parallel to the $xz$-plane at an angle $\phi$ relative to the $z$-axis (see Figure \figref{fig:figure2}{a}, \textit{Prism Perspective} coordinate axis).\newline
\indent Let:
\begin{align}
	n_1 = 1,\ 
	n_2 = 2 \quad\text{and}\quad\ 
	\mathbf{n} = \frac{1}{\sqrt{2}}\begin{pmatrix} 0\\ -1\\ 1 \end{pmatrix},\ 
	\mathbf{k}_{in} = \begin{pmatrix} \cos \phi\\ 0\\ \sin \phi \end{pmatrix}\implies\
	\mathbf{n}\cdot\mathbf{k}_{in} = \frac{1}{\sqrt{2}} \sin \phi.\nonumber
\end{align}
\noindent Inserting this into the vectorial equation $\mathbf{k}_{out}\left(\mathbf{n},\mathbf{k}_{in}\right)$ gives:
\begin{align}
\mathbf{k}_{out}=\left[\sqrt{1-\left(\frac{1}{2}\right)^2\left(1-\left( \frac{1}{\sqrt{2}} \sin \phi\right)^2\right)}-\left(\frac{1}{2}\right) \left(\frac{1}{\sqrt{2}} \sin \phi\right)\right]\cdot \frac{1}{\sqrt{2}}\begin{pmatrix} 0\\ -1\\ 1 \end{pmatrix} +\left(\frac{1}{2}\right) \begin{pmatrix} \cos \phi\\ 0\\ \sin \phi \end{pmatrix}.
\end{align}
\noindent Simplifying the equation further yields:
\begin{align}
	\mathbf{k}_{out} = \frac{1}{4}\left[\sqrt{6+\sin^2 \phi} - \sin \phi\right]\begin{pmatrix} 0\\ -1\\ 1 \end{pmatrix}+\frac{1}{2}\begin{pmatrix} \cos \phi\\ 0\\ \sin \phi \end{pmatrix}.\nonumber
\end{align}
In our laboratory coordinate system (c.f. \figref{fig:figure2}{a}, Prism Perspective) the x-coordinate can be expressed as $x = \cos \phi$, thus, $\phi = \arccos x$. We can then write $\mathbf{k}_{out}$ as a function of $x$. It is useful to calculate $\sin\left(\arccos x\right) = \sqrt{1-x^2}$.
\begin{align}
	\mathbf{k}_{out} = \frac{1}{4}\left[\sqrt{7-x^2} - \sqrt{1-x^2}\right]\begin{pmatrix} 0\\ -1\\ 1 \end{pmatrix}+\frac{1}{2}\begin{pmatrix} x\\ 0\\ \sqrt{1-x^2} \end{pmatrix}\nonumber	
\end{align}
\noindent To investigate what curve $\mathbf{k}_{out}$ is tracing out in the $xy$-plane, we can solely focus on the $y$-coordinate as a function of $x$.
\begin{align}
	Arc(x) = y\left(x\right) =  -\frac{1}{4}\left(\sqrt{7-x^2} - \sqrt{1-x^2}\right).\nonumber
\end{align}
\noindent The traced-out arc consists of the difference between two ellipses.
\begin{figure}[H]
	\centering
	\includegraphics[keepaspectratio]{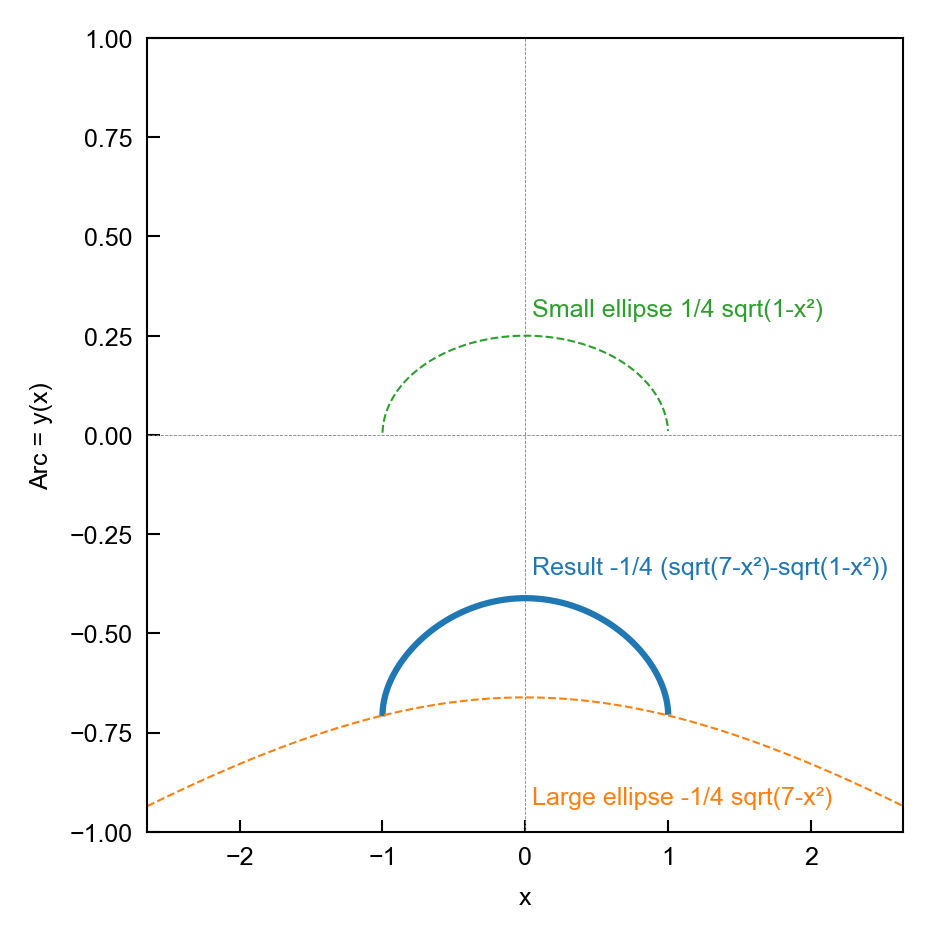}
	\caption{Resulting arc (blue) that is the difference of a large ellipse (orange) and a small ellipse (green).}
\end{figure}
\noindent We plotted $Arc(x)$ (blue) in the figure above to give an example to the reader.

\vfill
\newpage

\onecolumngrid

\subsection{Simulation principle and projections}
\begin{figure*}[t!]
	\centering
	\includegraphics[keepaspectratio]{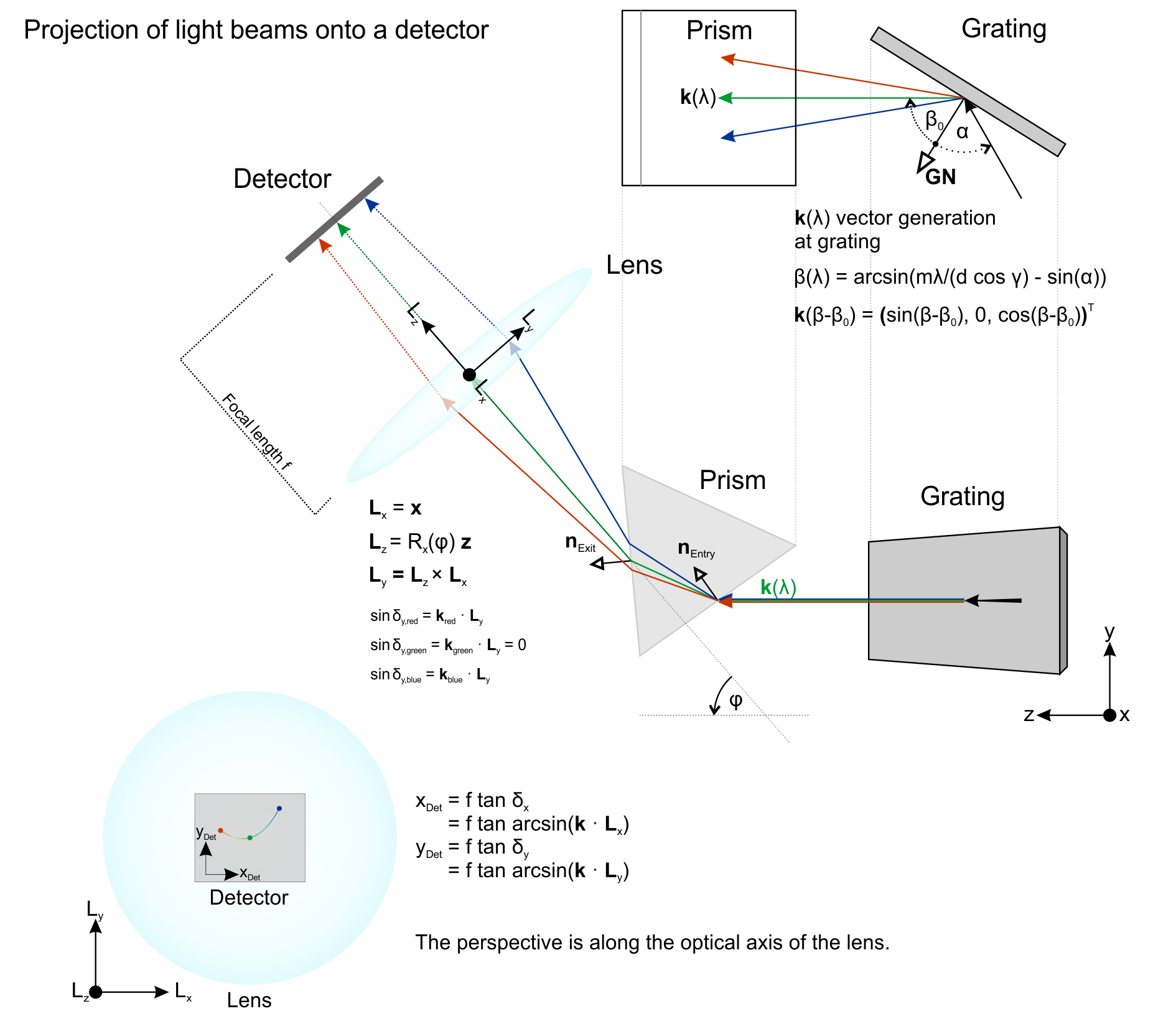}
	\caption{Modelling approach of \'{E}chelle spectrometers.}
	\label{fig:suppfigure3}
\end{figure*}
\noindent We carry our simulations out mostly by using Python3 (https://www.python.org), with the packages astropy, numpy, scipy, and matplotlib.
The procedure is as follows:
\begin{itemize}
	\item Define a laboratory coordinate system $\left\{\mathbf{x},\mathbf{y},\mathbf{z}\right\}$
	\item Define the normal vectors of the prism $\left\{\mathbf{n}_{EntryFacet}, \mathbf{n}_{ExitFacet}\right\}$ in beam direction.
	\item Define a lens coordinate system $\left\{\mathbf{L}_x,\mathbf{L}_y,\mathbf{L}_z\right\}$ for the image generation and projections $\mathbf{L}_z$ represents the optical axis.
	$\mathbf{L}_x$ and $\mathbf{L}_y$ help to project incoming beams within an orthogonal system. Example: 
	\begin{itemize}
		\item Rotate the $\mathbf{z}$ unit vector around $\mathbf{x}$ by $\phi$ via the rotation matrix $R_x(\phi)$ to generate $\mathbf{L}_z$
		\item Define $\mathbf{L}_x = \mathbf{x}$
		\item Calculate $\mathbf{L}_y = \mathbf{L}_z \times \mathbf{L}_x$
	\end{itemize}
	\item Define a wavelength range array $\lambda$[i] from 350 nm to 1150 nm
	\item Calculate an array of normalized $\mathbf{k}$ vectors {$\mathbf{k}_{in}$}[j] in the laboratory coordinate system $\left\{\mathbf{x},\mathbf{y},\mathbf{z}\right\}$ using the grating equation with an offset angle $\beta$0 (see figure) and the array $\lambda$[i] that will enter the next optical element. Here, the prism.
	\item For every $\mathbf{k}_{in}$ vector, calculate the transmitted $\mathbf{k}_{out}$ vector within the prism medium and repeat this for the light ray leaving the prism.
	Normalize the output vectors and compare for control.
	Snell’s law may not change the length. 
	\item Project the $\mathbf{k}$ vector after the prism onto the Lens coordinate axes.
	The resulting angles are needed to deduce the position on the detector when using the Lens equation $(x,y) = f \tan(\delta_{(x,y)})$, where $f$ is the focal length.
The detector can be defined collinear to the Lens coordinate system.
$\mathbf{x}_{Det} = \mathbf{L}_x$, $\mathbf{y}_{Det} = \mathbf{L}_y$.
Yet, if needed, it is possible to rotate the lens coordinate system around an arbitrary vector (e.g. around the optical axis).
SCIPY has very handy classes and methods to achieve this.
Now the projected $\mathbf{k}$ vectors will be rotated on the detector.
\end{itemize}
\newpage

\subsubsection{Python code example - Check QtYETI on github}
\begin{linenumbers}
	\centering
	\begin{Verbatim}
import numpy as np

@dataclass(init=True, repr=True)
  class Ray:
  lambda: float
  color: str
  direction: np.ndarray

def refraction_on_surface(beam_vector, surface_normal, refractive_index_1, refractive_index_2):
  """
  Snell's law in vector form
  --------------------------
  """
  vec_k = np.asarray(beam_vector)
  vec_sn = np.asarray(surface_normal)
  n1 = np.asarray(refractive_index_1,dtype=np.float64)
  n2 = np.asarray(refractive_index_2,dtype=np.float64)
  return (np.sqrt( 1 - ((n1/n2)**2) * (1 - np.dot(vec_sn, vec_k)**2) ) \ 
    - (n1/n2) * np.dot(vec_sn, vec_k)) * vec_sn + (n1/n2)*vec_k
	  
def grating_b(m, l, d, a, g):
  """
  Return the outgoing angle of a light ray for a given order m with a given wavelength
  and incident angles
  """
  return np.arcsin( ((m * l)/(d * np.cos(g))) - np.sin(a) )
	
for m in np.arange(-minimum_visible_order,-maximum_visible_order,-1):
  x_list = []
  y_list = []
  ray_list = []

  for lambda in spectrum:
    angle = grating_b(m,lambda,d,grating_a,grating_g)-(grating_b0)
    ray = Ray(lambda,wavelength_to_rgb(lambda),np.asarray([np.sin(angle),0,np.cos(angle)]))
    if (abs(angle) <= cam_fov_x_half):
      k_i = ray.direction
      k_in = k_i/np.linalg.norm(k_i)
      k_medium = \
        refraction_on_surface(k_in, prism_entry_surface_normal, n_air, sellmeier_n(ray.lambda)
      k_out = \
        refraction_on_surface(k_medium, prism_exit_surface_normal, sellmeier_n(ray.lambda), n_air)
      k_on = k_out/np.linalg.norm(k_out)
      
      # Project the resulting beam onto the optical axis of the lens
      # k_on • Lx = cos(..) → k_on • Lx = cos(90-a) → k_on • Lx = sin(a)
      a = np.arcsin(np.dot(k_on, Lx))
      g = np.arcsin(np.dot(k_on, Ly))
      x_pixel = x_pixels/2 + np.rint( (1/px_size) * focal_length*np.tan(a))
      y_pixel = y_pixels/2 + np.rint( (1/px_size) * focal_length*np.tan(-g))
      x_list.append(x_pixel)
      y_list.append(y_pixel)
\end{Verbatim}
\end{linenumbers}

\subsection{Thoughts on grating dispersion and order calibration}
\begin{figure*}[t!]
	\centering
	\includegraphics[keepaspectratio]{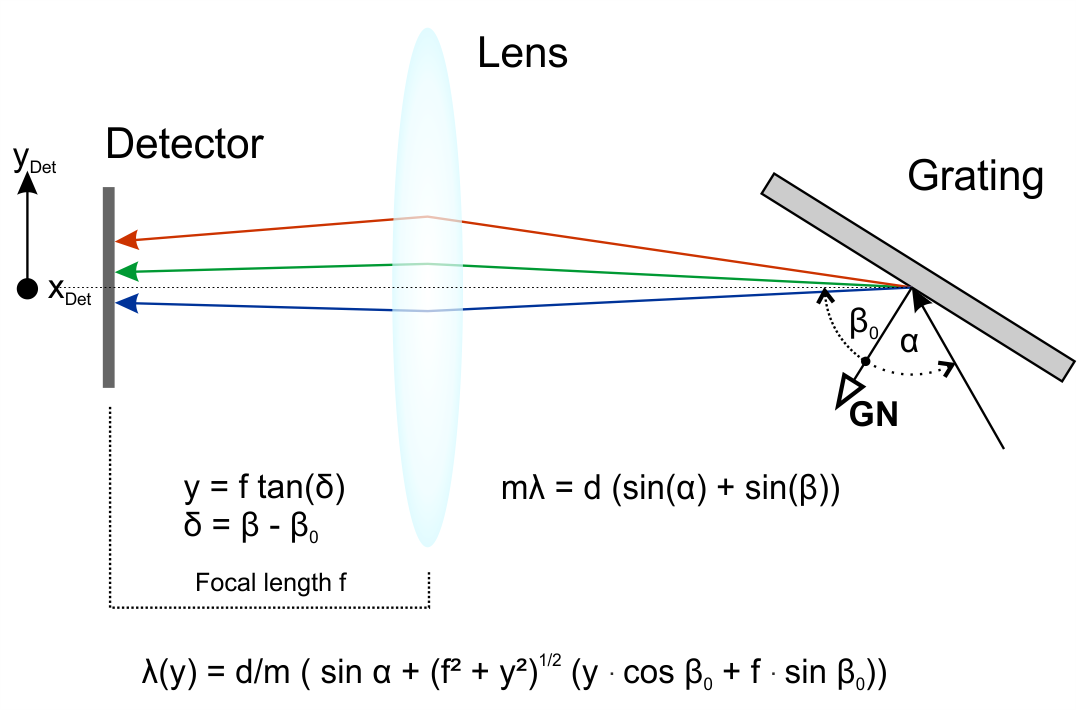}
	\caption{General dispersion formula for a reflective diffraction grating and a lens positioned at an angle $\beta_0$ relative to the grating normal $\mathbf{GN}$.}
	\label{fig:suppfigure4}
\end{figure*}
\noindent Prisms as cross-dispersers untangle the overlapping grating orders and colours as shown in the main text in Figure \ref{fig:figure1}.
They have no influence on the initial grating dispersion that happens in the plane of incidence.
Thus, we can simply treat the system as grating plus a lens.
Here, we see a lens with its optical axis being at an angle $\beta_0$ relative to the grating normal vector $\mathbf{GN}$.
By combining the grating and focusing equations, and using trigonometric identities, we obtain:
$$\lambda\left(y\right)=\frac{d \cdot \cos{\gamma}}{m}\left(\sin{\alpha}+\frac{1}{\sqrt{f^2+y^2}}\left(y\cdot\cos{\beta_0}+f\cdot\sin{\beta_0}\right)\right)$$
This equation can help to predict the in-plane position of spectral lines on a detector. Especially with negligible detector rotation. This is independent of any arcing from a oblique incidence on a prism.\vfill\eject
\newpage

\subsection{Overlap between measurement and predication}
\begin{figure*}[h!]
	\centering
	\includegraphics[width=1\textwidth,angle=0,origin=c]{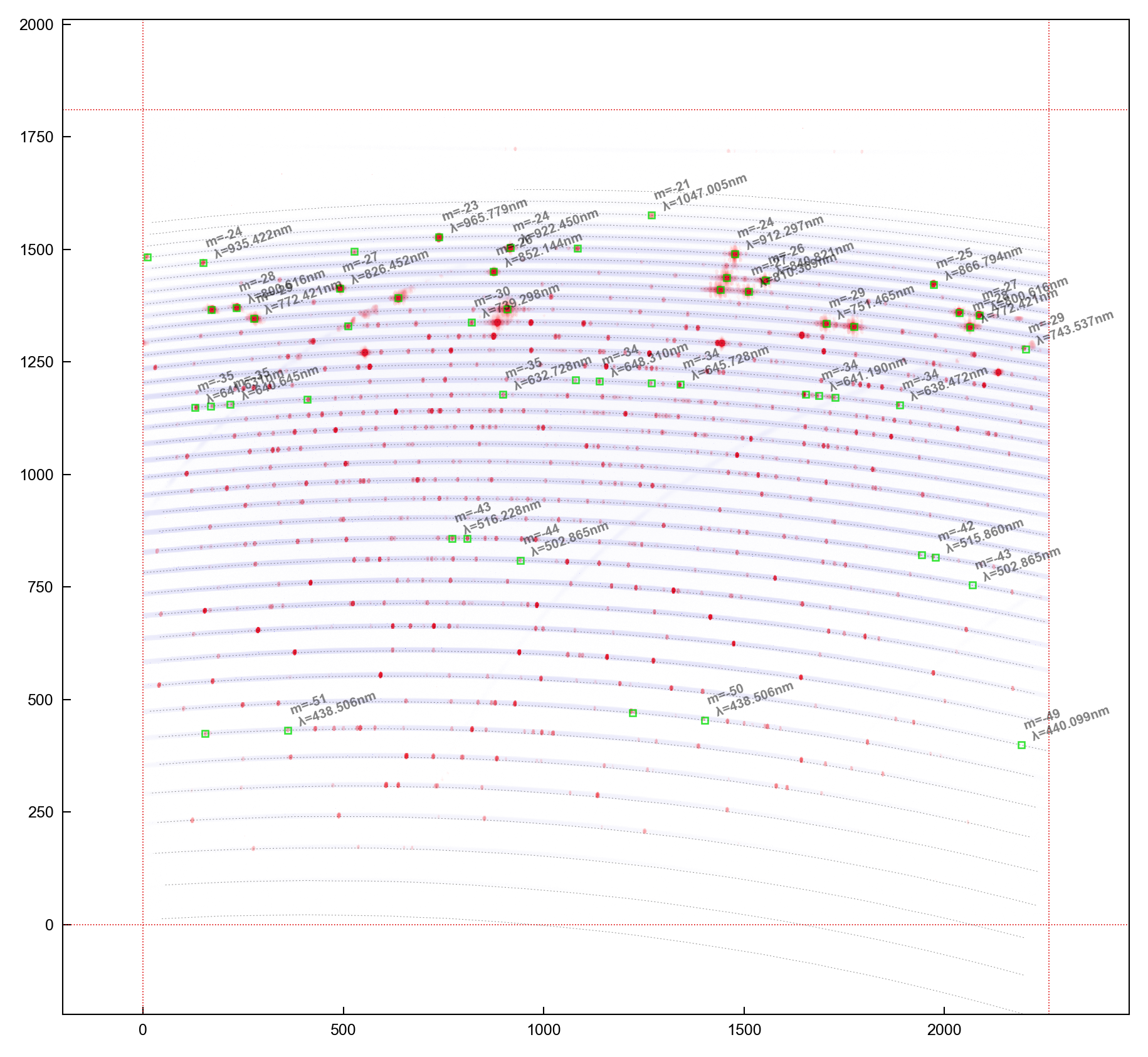}
	\caption{General dispersion.}
	\label{fig:suppfigure5}
\end{figure*}

\noindent Figure \ref{fig:suppfigure5} shows the overlap between a measured spectrogram (light blue, flat field image) and the predicted spectrogram (dotted, black). It is the magnified version \figref{fig:figure3}{a}.
The red dots are measured ThAr peaks within the visible orders. The green boxes denote simulated positions of randomly selected Ar and Th lines from NIST \cite{bib:NIST2023}.
The absolute orders and wavelengths in nanometres are added to every other green box for clarity.
It is evident, that the predicted and measured orders nicely overlap, showing that the algorithm works dependably.
For this to work ad-hoc one needs to understand the geometry of the spectrograph (incident angles) as well as the focal length of the objective in front of the detector.
Note: This spectrogram was taken on a home-built FLECHAS spectrograph by Bernd Bitnar. The built-up was supported by Karl-Heinz Wolf and
Klaus Vollmann.

\subsection{FLECHAS spectrometer}
\noindent The FLECHAS spectrometer created by the CAOS group (\url{https://spectroscopy.wordpress.com/2017/06/22/flechas-opto-mechanical-components/}) is a f/18 spectrometer with a resolution of $\delta\lambda/\lambda$ of 7000 – 11000 without an image slicer.
The parabolic mirror has a focal length of $f=\SI{444}{\milli\meter}$ resulting an a beam diameter of $\SI{25}{\milli\meter}$
Please refer to their excellent homepage containing years of experience and several spectrometers built by CAOS \cite{bib:Flechas2010}.

\end{document}